\documentclass[preprint,superscriptaddress,showpacs,preprintnumbers,amsmath,amssymb]{revtex4}
\usepackage{graphicx} 
\usepackage{dcolumn}  

\graphicspath{{ps}}

\begin{document}


\preprint{\vbox{ \hbox{   }
							   \hbox{Belle Preprint 2013-10} 
							   \hbox{KEK Preprint 2013-10}
}}

\title{ \quad\\[1.0cm] Evidence for Semileptonic \boldmath $ {B^- \to p\bar{p}\ell^- \bar{\nu}_\ell}$ Decays }


\begin{abstract}
We find evidence for the semileptonic baryonic decay
$B^-\to p\bar p\ell^-\bar\nu_\ell$ ($\ell=e,\mu$), based on a data
sample of 772 million $B\bar B$ pairs collected at the $\Upsilon(4S)$
resonance with the Belle detector at the KEKB asymmetric-energy
electron-positron collider. A neural-network based
hadronic $B$-meson tagging method is used in this study. The branching fraction of $B^-\to p\bar p\ell^-\bar\nu_\ell$ 
is measured to be $(5.8^{+2.4}_{-2.1}\textrm{(stat.)}\pm 0.9\textrm{(syst.)})\times 10^{-6}$ with a significance of 3.2$\sigma$, 
where lepton universality is assumed. We also estimate the corresponding upper limit: 
$\mathcal{B}(B^-\to p\bar p\ell^-\bar\nu_\ell) < 9.6\times 10^{-6}$ at the $90\%$ confidence level. This measurement helps constrain the baryonic transition form factor in $B$ decays.
\end{abstract}


\noaffiliation
\affiliation{University of the Basque Country UPV/EHU, 48080 Bilbao}
\affiliation{University of Bonn, 53115 Bonn}
\affiliation{Budker Institute of Nuclear Physics SB RAS and Novosibirsk State University, Novosibirsk 630090}
\affiliation{Faculty of Mathematics and Physics, Charles University, 121 16 Prague}
\affiliation{University of Cincinnati, Cincinnati, Ohio 45221}
\affiliation{Deutsches Elektronen--Synchrotron, 22607 Hamburg}
\affiliation{Justus-Liebig-Universit\"at Gie\ss{}en, 35392 Gie\ss{}en}
\affiliation{Gifu University, Gifu 501-1193}
\affiliation{Hanyang University, Seoul 133-791}
\affiliation{University of Hawaii, Honolulu, Hawaii 96822}
\affiliation{High Energy Accelerator Research Organization (KEK), Tsukuba 305-0801}
\affiliation{Ikerbasque, 48011 Bilbao}
\affiliation{Indian Institute of Technology Guwahati, Assam 781039}
\affiliation{Indian Institute of Technology Madras, Chennai 600036}
\affiliation{Institute of High Energy Physics, Chinese Academy of Sciences, Beijing 100049}
\affiliation{Institute of High Energy Physics, Vienna 1050}
\affiliation{Institute for High Energy Physics, Protvino 142281}
\affiliation{INFN - Sezione di Torino, 10125 Torino}
\affiliation{Institute for Theoretical and Experimental Physics, Moscow 117218}
\affiliation{J. Stefan Institute, 1000 Ljubljana}
\affiliation{Kanagawa University, Yokohama 221-8686}
\affiliation{Institut f\"ur Experimentelle Kernphysik, Karlsruher Institut f\"ur Technologie, 76131 Karlsruhe}
\affiliation{Korea Institute of Science and Technology Information, Daejeon 305-806}
\affiliation{Korea University, Seoul 136-713}
\affiliation{Kyungpook National University, Daegu 702-701}
\affiliation{\'Ecole Polytechnique F\'ed\'erale de Lausanne (EPFL), Lausanne 1015}
\affiliation{Faculty of Mathematics and Physics, University of Ljubljana, 1000 Ljubljana}
\affiliation{Luther College, Decorah, Iowa 52101}
\affiliation{University of Maribor, 2000 Maribor}
\affiliation{Max-Planck-Institut f\"ur Physik, 80805 M\"unchen}
\affiliation{School of Physics, University of Melbourne, Victoria 3010}
\affiliation{Moscow Physical Engineering Institute, Moscow 115409}
\affiliation{Moscow Institute of Physics and Technology, Moscow Region 141700}
\affiliation{Graduate School of Science, Nagoya University, Nagoya 464-8602}
\affiliation{Kobayashi-Maskawa Institute, Nagoya University, Nagoya 464-8602}
\affiliation{Nara Women's University, Nara 630-8506}
\affiliation{National Central University, Chung-li 32054}
\affiliation{National United University, Miao Li 36003}
\affiliation{Department of Physics, National Taiwan University, Taipei 10617}
\affiliation{H. Niewodniczanski Institute of Nuclear Physics, Krakow 31-342}
\affiliation{Nippon Dental University, Niigata 951-8580}
\affiliation{Niigata University, Niigata 950-2181}
\affiliation{University of Nova Gorica, 5000 Nova Gorica}
\affiliation{Osaka City University, Osaka 558-8585}
\affiliation{Pacific Northwest National Laboratory, Richland, Washington 99352}
\affiliation{Panjab University, Chandigarh 160014}
\affiliation{University of Pittsburgh, Pittsburgh, Pennsylvania 15260}
\affiliation{University of Science and Technology of China, Hefei 230026}
\affiliation{Seoul National University, Seoul 151-742}
\affiliation{Soongsil University, Seoul 156-743}
\affiliation{Sungkyunkwan University, Suwon 440-746}
\affiliation{School of Physics, University of Sydney, NSW 2006}
\affiliation{Tata Institute of Fundamental Research, Mumbai 400005}
\affiliation{Excellence Cluster Universe, Technische Universit\"at M\"unchen, 85748 Garching}
\affiliation{Toho University, Funabashi 274-8510}
\affiliation{Tohoku Gakuin University, Tagajo 985-8537}
\affiliation{Tohoku University, Sendai 980-8578}
\affiliation{Department of Physics, University of Tokyo, Tokyo 113-0033}
\affiliation{Tokyo Institute of Technology, Tokyo 152-8550}
\affiliation{Tokyo Metropolitan University, Tokyo 192-0397}
\affiliation{Tokyo University of Agriculture and Technology, Tokyo 184-8588}
\affiliation{CNP, Virginia Polytechnic Institute and State University, Blacksburg, Virginia 24061}
\affiliation{Wayne State University, Detroit, Michigan 48202}
\affiliation{Yamagata University, Yamagata 990-8560}
\affiliation{Yonsei University, Seoul 120-749}

\author{K.-J.~Tien}\affiliation{Department of Physics, National Taiwan University, Taipei 10617} 
 \author{M.-Z.~Wang}\affiliation{Department of Physics, National Taiwan University, Taipei 10617} 
  \author{I.~Adachi}\affiliation{High Energy Accelerator Research Organization (KEK), Tsukuba 305-0801} 
  \author{H.~Aihara}\affiliation{Department of Physics, University of Tokyo, Tokyo 113-0033} 
  \author{D.~M.~Asner}\affiliation{Pacific Northwest National Laboratory, Richland, Washington 99352} 
  \author{V.~Aulchenko}\affiliation{Budker Institute of Nuclear Physics SB RAS and Novosibirsk State University, Novosibirsk 630090} 
  \author{T.~Aushev}\affiliation{Institute for Theoretical and Experimental Physics, Moscow 117218} 
  \author{A.~M.~Bakich}\affiliation{School of Physics, University of Sydney, NSW 2006} 
  \author{A.~Bala}\affiliation{Panjab University, Chandigarh 160014} 
  \author{B.~Bhuyan}\affiliation{Indian Institute of Technology Guwahati, Assam 781039} 
% ...
\collaboration{The Belle Collaboration}
\noaffiliation

\maketitle


{\renewcommand{\thefootnote}{\fnsymbol{footnote}}}
\setcounter{footnote}{0}

\begin{figure}[htb]
\includegraphics[width=0.35\textwidth]{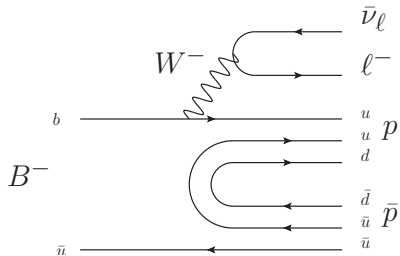}
\caption{Leading diagram for $B^-\to p\bar p\ell^-\bar\nu_\ell$ decay.}
\label{fig:ppln_fd}
\end{figure}

Measurements of charmless semileptonic $B$ decays play an important role in the determination of 
the fundamental parameter $|V_{ub}|$ of the Cabibbo-Kobayashi-Maskawa (CKM) matrix~\cite{ref:KM} in the Standard Model. However, all previous 
efforts have mainly been focused on $\bar{B} \to M l \bar{\nu_l}$~\cite{ref:Mlnu,ref:CC}, where $M$ stands for a charmless meson. 
There are no observations to date of semileptonic $B$ decays with a charmless
baryon-antibaryon pair in the final state. The most stringent upper limit to date has been set by the CLEO collaboration 
with $\mathcal{B}(B^-\to p\bar p e^-\bar\nu_e) < 5.2\times 10^{-3}$~\cite{ref:CLEO}. The corresponding decay diagram
is shown in Fig.~\ref{fig:ppln_fd}.

A theoretical investigation based on phenomenological arguments suggests
that the branching fraction of exclusive semileptonic $B$ decays to a baryon-antibaryon pair is only about $10^{-5} - 10^{-6}$~\cite{ref:HouSoni}, so sensitivity to such decays with the current data sets accumulated at the
$B$-factories is marginal. In fact, there have been no final states with charmed baryons to date in semileptonic $B$ decays. 
The \textsc{BaBar} collaboration only reported an upper 
limit of $\mathcal{B}(\bar{B} \to \Lambda^+_c X \ell^-\bar\nu_\ell)/\mathcal{B}(\bar{B} \to \Lambda^+_c X) < 3.5\%$ ~\cite{ref:BABAR}
at the 90\% confidence level (C.L.). 

A recent paper~\cite{ref:GengHsiao} used experimental inputs [8-12] to estimate the $B$ to baryon-antibaryon transition 
form factors predicted an unexpectedly large branching fraction, $(1.04 \pm 0.38) \times 10^{-4}$, 
for  $B^-\to p\bar p\ell^-\bar\nu_\ell$ ($\ell=e,\mu$). This is at the same level as many known $\bar{B} \to M l \bar{\nu_l}$ decays such as 
$\bar{B} \to \pi l \bar{\nu_l}$~\cite{ref:PDG}. This meta-analysis triggered our direct experimental search,
whose results could be used to improve the theoretical understanding of baryonic $B$ decays, 
if the predicted branching fraction is confirmed, 
many similar decays will become available and,
with improved theoretical understanding, they will
be helpful in determining $|V_{ub}|$ in future.

In this study, we use the full data set
of $772 \times 10^6\ B\bar{B}$ pairs collected at the $\Upsilon(4S)$ resonance
with the Belle detector~\cite{ref:Belle} at the KEKB asymmetric-energy
$e^+e^-$ (3.5 on 8~GeV) collider~\cite{ref:KEKB}.  
The Belle detector is a large-solid-angle magnetic
spectrometer that consists of a silicon vertex detector (SVD),
a 50-layer central drift chamber (CDC), an array of
aerogel threshold Cherenkov counters (ACC),  
a barrel-like arrangement of time-of-flight
scintillation counters (TOF), and an electromagnetic calorimeter
comprised of CsI(Tl) crystals (ECL) located inside 
a superconducting solenoid coil that provides a 1.5~T
magnetic field.  An iron flux return located outside of
the coil is instrumented to detect $K_L^0$ mesons and to identify
muons (KLM).  The detector
is described in detail elsewhere~\cite{ref:Belle}.

Monte Carlo (MC) event samples are simulated to evaluate signal efficiency, 
optimize selection criteria and determine the shapes for signal and 
background distributions in our analysis. For the signal decays, three million events are 
generated for each final state lepton flavour of electron or muon. 
The MC simulation takes into
account the experimental conditions pertaining to different running periods of the Belle experiment and the accumulated integrated luminosity for each 
period. Several MC samples are used to estimate four categories of background: continuum 
$(e^+e^-\rightarrow q\bar{q}$, where $q=u,d,s,c) $, $B\bar{B}$ (modelling $b \rightarrow c$ transitions), rare $B$ decays and charmless semileptonic $B$ decays 
($b \rightarrow u\ell \nu$ transitions), corresponding to 5, 5, 50 and 20 times the integrated luminosity of data, respectively. All MC samples are generated 
using the EvtGen~\cite{ref:EvtGen} package, and detector simulation 
is performed using GEANT~\cite{ref:GEANT}. 
Previous studies of similar baryonic $B$ decays, \textit{viz.}
$B^- \rightarrow p \bar{p} \pi^-$~\cite{ref:B_ppkpi}, 
$B^- \rightarrow p \bar{p} K^-$~\cite{ref:B_ppkpi,ref:B_ppkBABAR}, and $B^- \rightarrow p \bar{p} K^{*-}$~\cite{ref:B_ppkst},
found that the proton-antiproton mass distributions have 
low mass enhancements near threshold. 
We therefore assume that the $p\bar{p}$ pairs have an invariant mass distribution centred at 2.2 GeV$/c^2$ with a width of about 0.2 GeV$/c^2$. 

We use the hadronic-tag $B$ reconstruction method to study $B$ 
decays with a neutrino in the final state. 
Since the $\Upsilon(4S)$ decays predominantly into $B\bar{B}$~\cite{ref:PDG}, 
we fully reconstruct one $B$ meson with selected fully-hadronic charmed final states, 
called $B_\textrm{tag}$.
The NeuroBayes algorithm~\cite{ref:fullrecon} is used to provide an assessment for the quality of $B_\textrm{tag}$ reconstruction. A total of 615 exclusive
charged $B$ hadronic decay channels are considered in the
NeuroBayes neural network to reconstruct $B_\textrm{tag}$ candidates. 
We reconstruct signal $B$ candidates, called $B_\textrm{cand}$, from the remaining particles in the event. 
These candidates are reconstructed using final states consisting of three charged particles: one proton, one antiproton and one electron or muon. 
To identify the neutrino, we define the missing mass squared as 
\begin{equation}
M^{2}_\text{miss} = E^{2}_\text{miss}/c^4 - |{\vec{p}}_\text{miss}/c|^{2},
\end{equation}
where $E_\text{miss}$ and ${\vec{p}}_\text{miss}$ are 
the energy and momentum component of the
four-vector $\textit{P}_\text{miss} = \textit{P}_{e^+} + \textit{P}_{e^-}- \textit{P}_{B\textrm{tag}} - \textit{P}_{B\textrm{cand}}$ in the laboratory frame. In this study, we
accept events whose missing mass is in the range $-1$ GeV$^2/c^4 <M^2_\text{miss}<3 $ GeV$^2/c^4$.

We ensure that tracks used for $B_\textrm{cand}$ reconstruction have not been used in the $B_\textrm{tag}$ reconstruction. 
In order to remove the secondary tracks generated by hadronic interactions with the detector material, we require $|dz| < 2.0\,\textrm{cm}$ and $dr < 0.4\,\textrm{cm}$, 
where $dz$ and $dr$ denote the distances  at the point of closest approach to the interaction point (IP) along the positron beam and in the plane transverse to this axis, respectively.  
To identify charged particles, all relevant information provided 
by the CDC, TOF and ACC is taken into account. For lepton identification, 
additional information is provided by the ECL and KLM. We define $\mathcal{L}_{p}$, $\mathcal{L}_{K}$, $\mathcal{L}_{\pi}$, $\mathcal{L}_{e}$ and $\mathcal{L}_{\mu}$ as likelihoods for a particle to be identified as a proton, kaon, pion, electron, and muon, respectively, and the likelihood ratios: $\mathcal{R}_{p/K}=\mathcal{L}_{p}/(\mathcal{L}_p+\mathcal{L}_K)$, $\mathcal{R}_{p/\pi}=\mathcal{L}_{p}/(\mathcal{L}_p+\mathcal{L}_\pi)$, $\mathcal{R}_e = \mathcal{L}_e/(\mathcal{L}_e+\mathcal{L}_\textrm{other})$ and $\mathcal{R}_\mu = \mathcal{L}_\mu/(\mathcal{L}_\mu+\mathcal{L}_\textrm{other})$.
For a track to be identified as a proton, it is required to satisfy the condition $\mathcal{R}_{p/K}>0.6$ and $\mathcal{R}_{p/\pi}>0.6$, 
and $\mathcal{R}_e$ and $\mathcal{R}_\mu$ must be less than $0.95$ for lepton rejection.
To identify lepton candidates, tracks with $\mathcal{R}_e >0.6$, $\mathcal{R}_\mu <0.95$ are regarded as electrons and those with $\mathcal{R}_\mu>0.9$, $\mathcal{R}_e<0.95$ 
as muons. 
In the kinematic region of interest, charged leptons are identified with an efficiency of about $90\%$, while the probability of misidentifying a pion as an electron (muon) is $0.25\%$ $(1.4\%)$.
The proton identification efficiency is about $95\%$, while the probability of misidentifying a kaon or a pion as a proton is less than $10\%$.
The momentum of an electron (muon) candidate in the laboratory frame must be greater than $300$ ($600$) MeV$/c$. 
The lepton charge must be opposite that of the $B_\text{tag}$.

Tag-side $B$ mesons are identified using the beam-energy-constrained mass,  $M_{\rm bc} \equiv \sqrt{E^{*2}_\text{beam}/c^4 - |\vec{p}^*_B/c|^2}$, 
and the energy difference, $\Delta E \equiv E^*_B - E^*_\text{beam}$, where $E^*_\text{beam}$ is the run-dependent beam energy, 
and $E^*_B$ and $\vec{p}^*_B$ are the reconstructed energy and momentum, respectively, of the $B_\textrm{tag}$ in the rest frame of the $\Upsilon(4S)$. 
We require that $M_{\rm bc}>5.27$ GeV$/c^2$ and $-0.15$ GeV$<\Delta E<0.1$ GeV to reject poorly reconstructed $B_\textrm{tag}$ candidates. 
The differences in event topology between the more spherical $B\bar{B}$ events and the dominant jet-like continuum background is used to suppress the latter. 
Here, the ratio of the second to zeroth Fox-Wolfram moments~\cite{ref:SFW}, the angle between the $B_\textrm{tag}$ direction and the thrust axis, and the angle 
between the $B_\textrm{tag}$ direction and the beam direction in the $\Upsilon (4S)$ rest frame are used to construct a NeuroBayes output value for continuum suppression ${o}^\textrm{cs}_\textrm{tag}$. The $B_\textrm{tag}$ with the largest value of ${o}^\textrm{cs}_\textrm{tag}$ within a given event is retained; 
we accept events satisfying $ \ln ({o}^\textrm{cs}_\textrm{tag})>-7 $ for $B^-\to p\bar pe^-\bar\nu_e$ and $ \ln({o}^\textrm{cs}_\textrm{tag})>-6 $ 
for $B^-\to p\bar p\mu^-\bar\nu_\mu$, according to the MC-determined selection optimization.

Since there can be more than one $B_\textrm{cand}$ in an event, we select the candidate with the smallest $\chi^2$ value obtained 
from a fit to the $B$ vertex. The fraction of events with multiple candidates is estimated
from MC to be
$0.21\%$ for $B^-\to p\bar pe^-\bar\nu_e$ and $0.17\%$ for $B^-\to p\bar p\mu^-\bar\nu_\mu$. The overall signal efficiency obtained is $0.279\%$ for $B^-\to p\bar pe^-\bar\nu_e$ and $0.222\%$ for $B^-\to p\bar p\mu^-\bar\nu_\mu$.
Since the reconstruction efficiency may differ between data and MC,  
we correct these efficiency estimates based on control sample studies. 
For proton and lepton identification, 
we use $\Lambda \rightarrow p\pi^-$ and $\gamma\gamma\rightarrow\ell^+\ell^-$ samples, respectively. 
The corrections are about $-4.4\%$ and $-3.1\%$ for $B^-\to p\bar pe^-\bar\nu_e$ and $-5.7\%$ and $-1.7\%$ for $B^-\to p\bar p\mu^-\bar\nu_\mu$. 
For the $B_\textrm{tag}$ reconstruction efficiency, we use $B^-\rightarrow X^0_c\ell^- \bar{\nu}_\ell$ samples, where $X^0_c$ denotes a meson containing a $c$ quark, and 
estimate correction factors of $-14.8\%$ for $B^-\to p\bar pe^-\bar\nu_e$ and $-16.4\%$ for $B^-\to p\bar p\mu^-\bar\nu_\mu$. 
Applying these corrections, the signal efficiency in data is estimated to be $(0.220\pm 0.011)\%$ for $B^-\to p\bar pe^-\bar\nu_e$ 
and $(0.172\pm 0.008)\%$ for $B^-\to p\bar p\mu^-\bar\nu_\mu$.

We perform a one-dimensional extended unbinned likelihood fit that maximizes the function
\begin{eqnarray}
\mathcal{L} &=&\frac{e^{-(N_\textrm{sig}+N_\textrm{bkg})}}{N!} \prod_{i=1}^N [N_\textrm{sig} P_\textrm{sig}({M^2_\textrm{miss}}^i)+N_\textrm{bkg} P_\textrm{bkg}({M^2_\textrm{miss}}^i)],
\end{eqnarray}
where $i$ is the event index, $N_\textrm{sig}$ and $N_\textrm{bkg}$ denote the fitted yields of signal 
and background, and $P_\textrm{sig}$ and $P_\textrm{bkg}$ denote the probability density functions (PDFs) 
in our signal extraction model. 
We use three Gaussian functions to describe $P_\textrm{sig}$ for $B^-\rightarrow p\bar{p}e^- \bar{\nu}_e$ 
and for $B^-\rightarrow p\bar{p}\mu^- \bar{\nu}_\mu$. For background, since no peak is present near the signal region, we combine 
both continuum and $B$ decays backgrounds to form one PDF. We use a normalized second-order Chebyshev polynomial function 
to represent $P_\textrm{bkg}$ for each mode. 
The shape of the signal PDF is determined from the MC simulation, while the shape of the background is floated. 
The rare $B$ decay and $b \rightarrow u\ell \nu$ backgrounds are not included in the fit, because less than 0.1 events are expected to be found on average in the fitting region.  

\begin{figure}[htbp]
\includegraphics[width=0.3\textwidth]{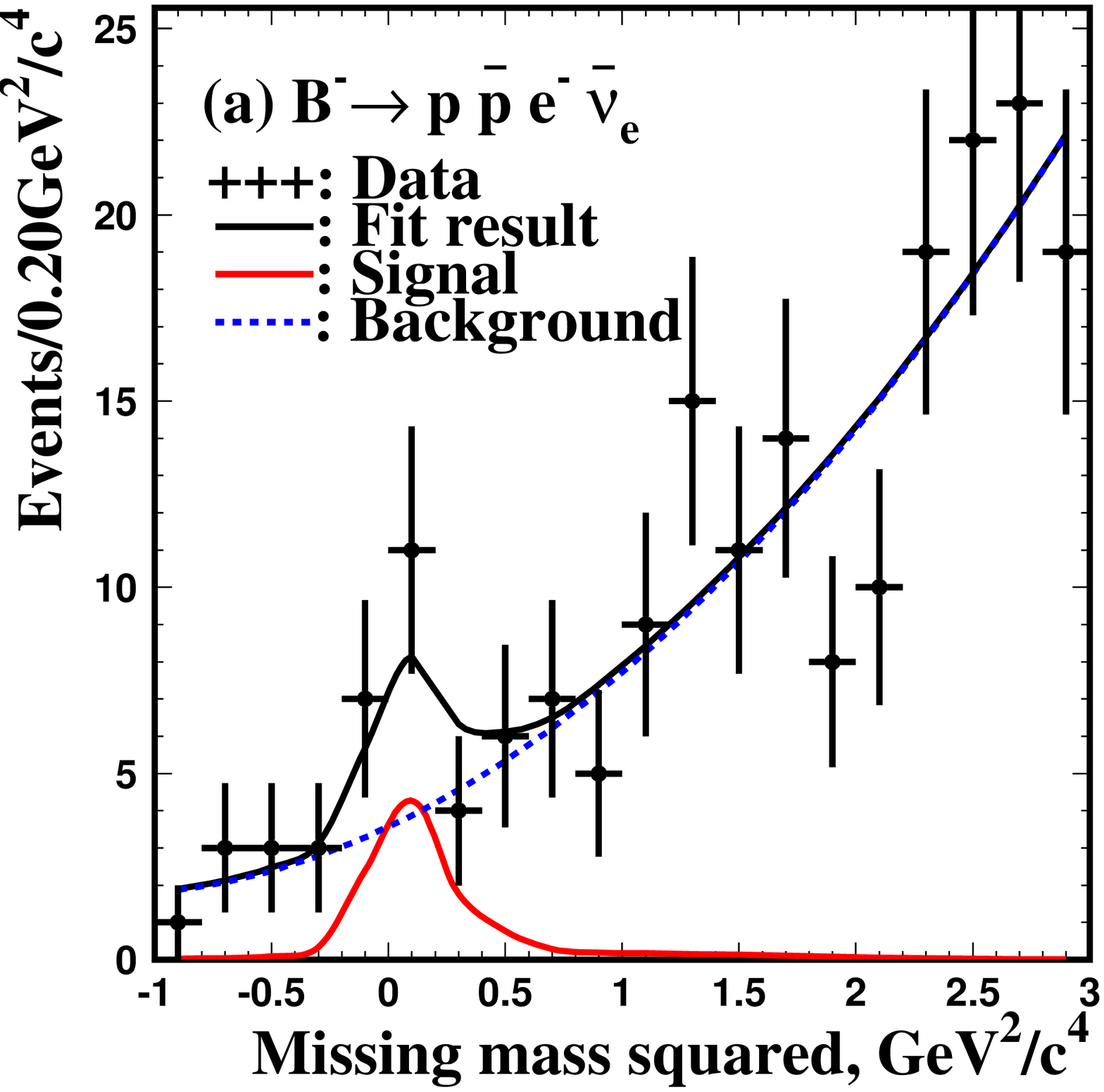}
\includegraphics[width=0.3\textwidth]{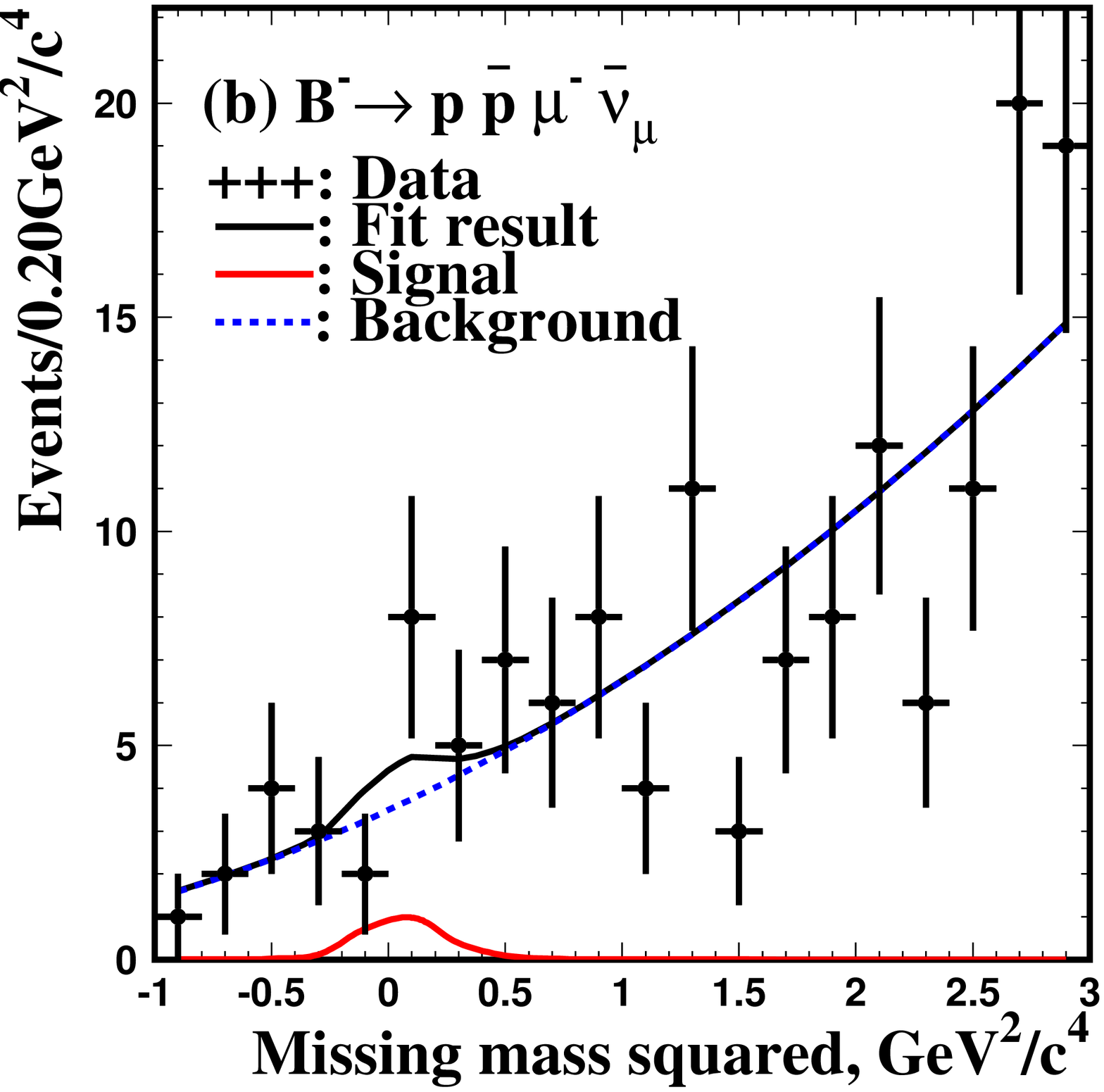}
\includegraphics[width=0.3\textwidth]{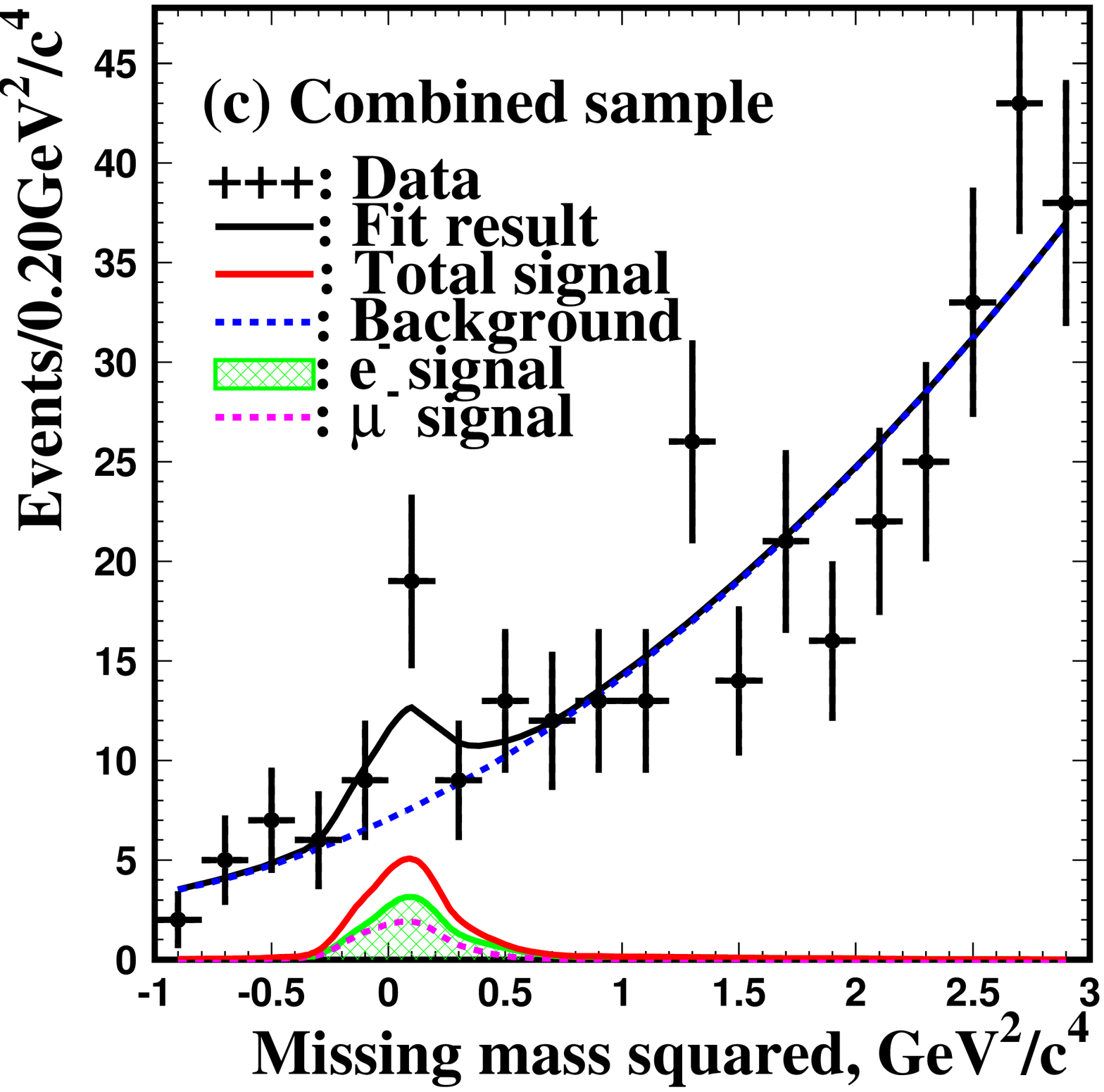}
\caption{Fitted missing mass squared distributions for 
(a) $B^-\to p\bar pe^-\bar\nu_e$, (b) $B^-\to p\bar p\mu^-\bar\nu_\mu$ and (c) the combined fit. 
Points with error bars represent data, 
while the curves denote various components of the fit: 
signal (solid red), total background (dashed blue), and the sum of all components (solid black).
The hatched green area denotes the signal fit component from $B^-\to p\bar pe^-\bar\nu_e$ and the dashed purple curve that from $B^-\to p\bar p\mu^-\bar\nu_\mu$.}
\label{fig:fitresult}
\end{figure}

The fit results are shown in Fig.~\ref{fig:fitresult}. 
We determine the fit significance in terms of $\sigma$, the standard deviation of a normal distribution, with
$\sqrt{-2\ln\left(\mathcal{L}_{0}/\mathcal{L}_\textrm{max}\right)}$, 
where $\mathcal{L}_0$ and $\mathcal{L}_\textrm{max}$ represent the maximum likelihood values from the fit with  
$N_\textrm{sig}$ set to zero, and with all parameters allowed to float, respectively. 
We also take into account the systematic effects from the signal decay model and PDF shape.
The significance 
is $3.0\sigma$ for $B^-\rightarrow p\bar{p}e^- \bar{\nu}_e$ 
and $1.3\sigma$ for $B^-\rightarrow p\bar{p}\mu^- \bar{\nu}_\mu$. 
Assuming lepton universality and equal branching fractions for $B^-\rightarrow p\bar{p}e^- \bar{\nu}_e$ 
and $B^-\rightarrow p\bar{p}\mu^- \bar{\nu}_\mu$, 
we obtain a combined fit result with a significance of $3.2\sigma$.

The systematic uncertainties on the branching fractions are summarised in Table~\ref{table:sys} and described below. 
Correlated (uncorrelated) errors are added linearly (in quadrature). 
Each systematic uncertainty for the combined fit is conservatively considered to be 
the larger of the uncertainties for $B^-\to p\bar pe^-\bar\nu_e$ and $B^-\to p\bar p\mu^-\bar\nu_\mu$, except for the fitting region uncertainty.

\begin{table}[hbtp]
\caption{Systematic uncertainties on the branching fractions, in percent.}
\label{table:sys}
\begin{tabular}
{@{\hspace{0.1cm}}l@{\hspace{0.1cm}} @{\hspace{0.5cm}}c@{\hspace{0.1cm}}
 @{\hspace{0.1cm}}c@{\hspace{0.1cm}} @{\hspace{0.5cm}}c@{\hspace{0.1cm}}}
\hline \hline
Source & {$p\bar{p}e^- \bar{\nu}_e$} & {$p\bar{p}\mu^-\bar{\nu}_\mu$} &Combined\\\hline
Track reconstruction & 1.1 & 1.1 & 1.1 \\
Proton identification & 0.7 & 0.8 & 0.8 \\
Lepton identification & 2.3 & 1.1 & 2.3\\
Tag calibration & 4.3 & 4.3 & 4.3\\
Number of $B\bar{B}$ events & 1.4 & 1.4 & 1.4\\
Signal Decay Model & 3.6 & 12 & 12\\
PDF Shape & 2.1 & 2.8 & 2.8\\
Fitting Region & 3.9 & 18 & 6.1\\\hline
Summary & 6.7 & 23 & 15\\\hline\hline
\end{tabular}
\end{table}

The systematic uncertainty due to charged-track reconstruction is estimated to be $0.35\%$ per track, using partially reconstructed $D^{*+}\rightarrow D^0(\pi^+\pi^-\pi^0)\pi^+$ decays. 
We estimate the uncertainty due to proton and lepton identification using the $\Lambda \rightarrow p\pi^-$ and $\gamma\gamma\rightarrow\ell^+\ell^-$ samples, respectively.
For tag calibration, the uncertainties are estimated to be $4.3\%$ for each of the two modes, using the $B^-\rightarrow X^0_c\ell^- \bar{\nu}_\ell$ sample.
The uncertainty due to the error on the total number of $B\bar{B}$ pairs is $1.4\%$. 
The uncertainty due to the signal MC modeling of the $p\bar{p}$ mass threshold enhancement is obtained by comparing the efficiency
difference between signal MC and the phase space decay model. 
The uncertainties due to the signal PDF shape are studied by varying each Gaussian parameter by $\pm 1\sigma$ and observing the yield difference.
Finally, the upper bound chosen for the fitting region which has a large effect on the fit results is varied from 2 to 4 GeV$^2/c^4$ with a step size of 0.2 GeV$^2/c^4$; we take one standard deviation of 
the ensemble of obtained fit results to estimate the uncertainty. 
These are conservative estimates as the statistical uncertainty is also included.

In addition to quoting branching fractions, we also estimate the corresponding upper limits at the $90\%$ confidence level by 
finding the value of $N$ that satisfies: 
\begin{equation}
\int^N_0 \mathcal{L}(n)dn = 0.9 \int^\infty_0\mathcal{L}(n)dn,
\end{equation}
where $\mathcal{L}(n)$ denotes the likelihood of the fit result and $n$ is the number of signal events. 
The systematic uncertainties are taken into account by replacing $\mathcal{L}(n)$ with a smeared likelihood function:
\begin{equation}
\mathcal{L}(n)=
\int^\infty_{-\infty}\mathcal{L}(n')
\frac{e^{-(n-n')^2/2\sigma^2_\textrm{syst.}}}{\sqrt{2\pi}\sigma_\textrm{syst.}}dn',
\end{equation}
where $\sigma_\textrm{syst.}$ is the systematic uncertainty of the associated signal yield $n'$.
 
Table~\ref{table:BRestimate} summarizes our results. The upper limits include systematic uncertainties. 

\begin{table}[htb]
\caption{Measured results and upper limits for the branching fractions ($\mathcal{B}$), where systematic uncertainties are taken into account.}
\label{table:BRestimate}
\extrarowheight=2.pt
\begin{tabular}
{@{\hspace{0.1cm}}l@{\hspace{0.1cm}} @{\hspace{0.5cm}}c@{\hspace{0.1cm}}
 @{\hspace{0.1cm}}c@{\hspace{0.1cm}} @{\hspace{0.5cm}}c@{\hspace{0.1cm}}}
\hline\hline
Mode & $\mathcal{B}$ $(10^{-6})$ & U.L. $(10^{-6})$\\\hline
$B^-\rightarrow p\bar{p}e^- \bar{\nu}_e$ & $8.2$ $^{+3.7}_{-3.2}\pm 0.6 $ & $13.8 $\\
$B^-\rightarrow p\bar{p}\mu^- \bar{\nu}_\mu$ & $3.1$ $^{+3.1}_{-2.4}\pm 0.7$ & $8.5$\\
Combined sample & $5.8$ $^{+2.4}_{-2.1}\pm 0.9$ & 9.6 \\\hline\hline

\end{tabular}
\end{table}

In conclusion, we have performed a search for the four-body semileptonic baryonic $B$ decay $B^-\to p\bar p\ell^-\bar\nu_\ell$ ($\ell=e,\mu$) 
using a neural-network based hadronic $B$ tagging method.
We find evidence for a signal with a significance of $3.2\sigma$ and a 
branching fraction of $(5.8^{+2.4}_{-2.1}\textrm{(stat.)}\pm 0.9\textrm{(syst.)})\times 10^{-6}$. 
This measurement is consistent with the theoretical investigation in Ref.~\cite{ref:HouSoni}.
As the statistical significance of our reported evidence is marginal, we also set an upper limit on the branching fraction: $\mathcal{B}(B^-\to p\bar p\ell^-\bar\nu_\ell) < 9.6\times 10^{-6}$ ($90\%$ C.L.). 
Our result is clearly lower than the recent meta-analysis expectation of $\sim 10^{-4}$~\cite{ref:GengHsiao}. 
It will be interesting to investigate the theoretical modeling of the baryonic transition form factors in $B$ decays in light of this new information.
With the proposed next generation $B$-factories, such semileptonic baryonic $B$ decays can be studied precisely and future results may be useful in further constraining the corresponding CKM matrix elements.

We thank the KEKB group for excellent operation of the
accelerator; the KEK cryogenics group for efficient solenoid
operations; and the KEK computer group, the NII, and 
PNNL/EMSL for valuable computing and SINET4 network support.  
We acknowledge support from MEXT, JSPS and Nagoya's TLPRC (Japan);
ARC and DIISR (Australia); NSFC (China); MSMT (Czechia);
CZF, DFG, and VS (Germany);
DST (India); INFN (Italy); MEST, NRF, GSDC of KISTI, and WCU (Korea); 
MNiSW and NCN (Poland); MES and RFAAE (Russia); ARRS (Slovenia);
IKERBASQUE and UPV/EHU (Spain); 
SNSF (Switzerland); NSC and MOE (Taiwan); and DOE and NSF (USA).


\begin{thebibliography}{99}
\bibitem{ref:CC}
Throughout this paper, the inclusion of the charge-conjugate mode decay is implied unless otherwise stated.
\end{thebibliography}
\end{document}